\documentclass{iau}
\usepackage{graphicx} 

\title[IAUS291.~~\it{Chandra} Observations of Black Widow Pulsars] %% short title %%
{\emph{Chandra} Observations of Black Widow Pulsars} %% full title %%

\author[P. Gentile, M. McLaughlin, M. Roberts]  %% short author list %%
{P. Gentile$^1$, M. McLaughlin$^1$,  M. Roberts$^2$, F. Camilo$^3$, J. Hessels$^4$, M. Kerr$^5$, S. Ransom$^6$, P. Ray$^7$, I. Stairs$^8$}

\affiliation{$^1$ Dept. of Physics, West Virginia University, Morgantown, WV 26506, USA \\
$^2$ Eureka ScientiÞc, Inc., Oakland, California 94602, USA\\
$^3$ Columbia Astrophysics Laboratory, Columbia University, New York, NY 10027, USA\\
$^4$ ASTRON, Postbus 2, 7990 AA Dwingeloo, The Netherlands\\
$^5$ W. W. Hansen Experimental Physics Laboratory, Kavli Institute for Particle Astrophysics and Cosmology, Department of Physics and SLAC National Accelerator Laboratory, Stanford University, Stanford, CA 94305, USA\\
$^6$ NRAO, 520 Edgemont Road, Charlottesville, Virginia 22093, USA\\
$^7$ Space Science Division, Naval Research Laboratory, Washington, DC 20375-5352, USA\\
$^8$ Department of Physics and Astronomy, University of British Columbia, 6224 Agricultural Road Vancouver, BC V6T 1Z1 Canada\\}

\pubyear{2012}
\volume{291}  %% insert here IAU Symposium No.
\jname{\mbox{Neutron Stars and Pulsars: Challenges and Opportunities after 80 years}}
\editors{J. van Leeuwen, ed.} 
\begin{document}

\maketitle

%% -- Abstract ----------------------------------
\begin{abstract}
We describe the first X-ray observations of binary millisecond pulsars PSRs J0023+0923, J1810+1744, J2215+5135, and J2256$-$1024. All four are  {\it Fermi} gamma-ray sources  and three are `black-widow' pulsars, with companions of mass $<0.1~M_{\odot}$.  Data were taken using the \textit{Chandra} X-Ray Observatory and covered a full binary orbit for each pulsar. Two pulsars, PSRs J2215+5135 and J2256$-$1024, show significant orbital variability and X-ray flux minima at the times of  eclipses observed at radio wavelengths. This phenomenon is consistent with intrabinary shock emission characteristic of black-widow pulsars. The other two pulsars, PSRs J0023+0923 and J1810+1744, do not demonstrate significant variability, but are fainter than the other two sources. Spectral fits yield power-law indices that range from 1.4 to 2.3 and blackbody temperatures in the hundreds of eV. The spectrum for PSR J2215+5135 shows a significant hard X-ray component (41\% of counts are above 2 keV), which is additional evidence for the presence of intrabinary shock emission.
\end{abstract}

% add below any authors, subjects and objects for indexing 
%   add more lines if necessary
%   but leave all lines commented out
%\index[author]{LastName1, Initials|textbf}
%\index[author]{LastName2, Initials|textbf}
%\index[subject]{Keyword1}
%\index[subject]{Keyword2}
%\index[object]{Object1}
%\index[object]{Object2}

\section{Introduction}

Of the roughly 2000 radio pulsars known today, about 10\% are millisecond pulsars (MSPs), old neutron stars which have been spun up, or `recycled', through accretion of material from a companion. Many details of this recycling process remain unknown, but it is clear that most known MSPs have degenerate white dwarf companions with masses between 0.1 and 0.4~$M_\odot$. However, some (up to 25\%) of MSPs are isolated. The process through which these MSPs were formed is unclear. One attractive theory is that the companions of isolated MSPs were ablated through energetic particles, X-rays, and/or gamma-rays produced in the intrabinary shock between the pulsar wind and that of the companion.

 The first pulsar showing evidence for the ablation process was the black-widow pulsar PSR B1957+20, which has an extremely low mass companion (0.020 $M_\odot$), shows radio eclipses due to absorption in the wind of the companion and dramatic pulse delays around the time of eclipse due to propagation through the wind. A 2002 ACIS-S observation (OBSID 1911) revealed unresolved synchrotron emission that is modulated throughout the orbit,  with roughly half of the emission due to the interaction of the pulsar and stellar winds and half from the pulsar itself. This observation showed a dip in the lightcurve no wider than one tenth of an orbit at an orbital phase of 0.25. (\cite{2003Sci...299.1372S})

The body of knowledge regarding black-widow pulsars is still lacking. We present analysis of rare observations of four pulsars with low mass companions, each observed for approximately one orbit.

% CUP work flow only accepts EPS -- not PDF, JPG, etc.
% \begin{figure}[b]
% \begin{center}
%  \includegraphics[width=3.4in]{YourFig.eps} 
%  \caption{Path of pre-solar grains from their stellar sources to the
%    laboratory.} 
%    \label{fig1}
% \end{center}
% \end{figure}

\section{Observations and Analysis}

\begin{table}[h!]
\begin{center}
\footnotesize
\begin{tabular}{ccccccccc}
\hline
\hline
\bf Name & \bf P$_{\bf \rm spin}$  & \bf DM  &   \bf P$_{\bf \rm orb}$ &  \bf M$_{\bf \rm c,min}$ & \bf T$_{\bf \rm obs}$ & \bf Cts & \bf kT & $\bf \Gamma$\\
            & (ms) & (pc/cm$^{3}$) & (hr) &   (M$_{\odot}$)  & (ks) &   & (eV) &    \\
\hline

J0023+0923 & 3.05 & 14.3 & 3.3 & 0.016 & 17 & 43 & 160$^{+80}_{-20}$ & 3.3$^{+0.5}_{-0.4}$  \\
J1810+1744 & 1.66 & 39.7 & 3.6 & 0.035 & 20 & 55 & -- & 2.3$^{+0.5}_{-0.6}$ \\
J2215+5135 & 2.61 & 69.2 & 4.2 & 0.22 & 17 & 133 & -- & 1.4$^{+0.3}_{-0.2}$\\
J2256$-$1024 & 2.29 & 13.8 & 5.1 & 0.030 & 20 & 141 & 170$^{+20}_{-20}$ & 1.5$^{+1.1}_{-1.3}$ \\
\hline
{\it B1957+20} & {\it 1.60} & {\it 29.1}  & {\it 9.1} & {\it 0.020} & {\it 43} & {\it 370} & {\it -- } & {\it 1.9$^{+0.5}_{-0.5}$} \\
\hline
\end{tabular}
\end{center}
\caption{Timing and X-ray Properties of Four {\it Fermi}-Associated Radio MSPs. Timing properties are from 350-MHz observations with the GBT (\cite{2010HEAD...11.1609B}). PSR J0023+0923 was fit with both a blackbody model and a power law model (separately), while PSR J2256$-$1024 was fit with a combined blackbody and power law model. PSRs J1023+0038 and B1957+20 are shown for comparison.}
\label{table:table1}
\end{table}

We observed PSRs J0023+0923, J1810+1744, J2215+5135, and J2256$-$1024 for 15~ks, 22~ks, 19~ks, and 22~ks respectively. The data were taken using \textit{Chandra}'s ACIS-S mode and analyzed using \textit{Chandra}'s data analysis suite, CIAO (verson 4.2). 

\begin{figure}[h]
\begin{center}
\includegraphics[width=5.4in]{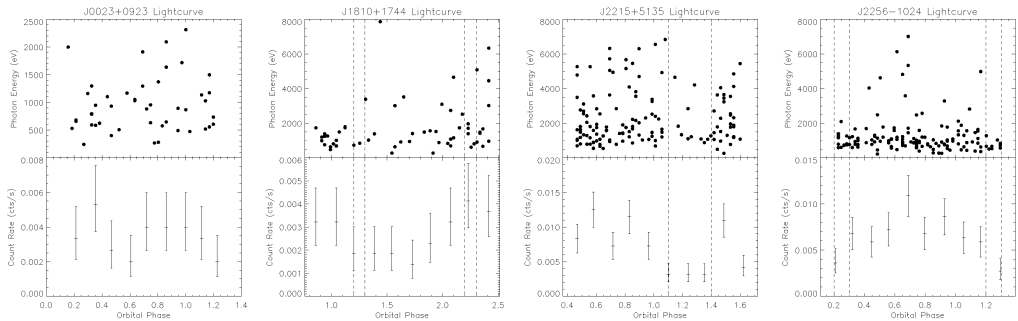}
\caption{Lightcurves for the four observed MSPs. Each bin corresponds to one tenth of an observation, and the dotted lines correspond to radio eclipse times. PSR J0023+0923 does not show a radio eclipse.  Note the absence of photons (specifically hard photons) during the radio eclipses of both PSRs J2215+5135 and J2256$-$1024.}
\end{center}
\end{figure}

Lightcurves were determined for each source using counts in the 0.2~keV to 8~keV range, as \textit{Chandra} has very little effective area outside of that range. The number of counts detected for each source ranged from 43 to 141 (Table 2). Each lightcurve was binned such that each bin represents one tenth of the observation. These lightcurves were then compared to uniform distributions using the $\chi^{2}$ test and Kolmogorov-Smirnov (K-S) test (\cite{kstest}) to determine their orbital variability.

\begin{figure}[h]
\begin{center}
\includegraphics[width=4in]{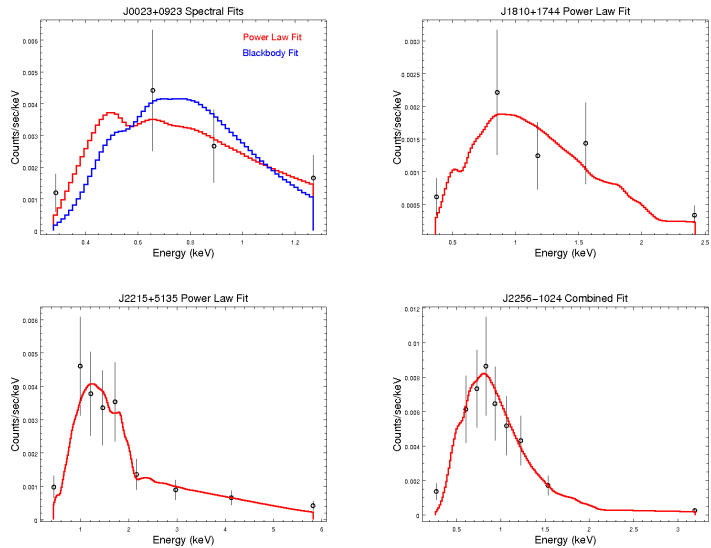}
\caption{Fitted spectra for PSRs J0023+0923, J1810+1744, J2215+5135, and J2256$-$1024, respectively (clockwise from top left). We were able to fit PSR J0023+0923 with both a blackbody model and a power law model (separately), while PSR J2256$-$1024 was fit with a combined blackbody and power law model. PSRs J1810+1744 and J2215+5135 could only be fit with a power law model.}
\end{center}
\end{figure}

Spectra were then analyzed using \textit{Chandra}'s spectral fitting platform, Sherpa. The data were then fitted over energies between 0.2 keV and 8 keV. Because of the small number of counts, we fixed $N_H$ at a constant value (with 10 free electrons per neutral Hydrogen) in all of the fits. Although in reality, the spectra for all of the sources will likely contain a blackbody component and a power-law component, we were not able to fit both of these components for PSRs J0023+0923, J1810+1744, and J2215+5135. Therefore, we provide the results of single-component fits for these pulsars. These fits are consistent with intrabinary shock emission, however, additional observations  are necessary to constrain both model components, as well as model the orbital geometry of these binary systems.

\end{document}